\documentclass[oldversion]{my_aa}  
\usepackage{graphicx,psfig,amsmath}

\newcommand{\bpicb}{$\beta$\,Pic\,b}
\newcommand{\bp}{$\beta$\,Pic}

\begin{document}
   \title{Is Beta\,Pic\,b the transiting planet of November 1981?}

   \author{
   A.~Lecavelier des Etangs\inst{1,2}
   \and
   A.~Vidal-Madjar\inst{1,2}       
 }
   

   \offprints{A.L. (\email{lecaveli@iap.fr})}

   \institute{
   CNRS, UMR 7095, 
   Institut d'Astrophysique de Paris, 
   98$^{\rm bis}$ boulevard Arago, F-75014 Paris, France
   \and
   UPMC Univ. Paris 6, UMR 7095, 
   Institut d'Astrophysique de Paris, 
   98$^{\rm bis}$ boulevard Arago, F-75014 Paris, France
   }
   
   \date{} 
 
  \abstract
{
   In 1981, $\beta$\,Pictoris showed strong and rapid 
   photometric variations that were 
   attributed to the transit of a giant comet or a planet orbiting at 
   several AUs. 
   Recently, a candidate planet 
   has been identified by imagery in the circumstellar disk of
   $\beta$\,Pictoris.
   This planet, named \bpicb , is observed at a projected distance of 8\,AU from 
   the central star. It is therefore 
   a plausible candidate for the photometric event observed in 1981. 
  The coincidence of the observed position of the planet in November 2003 
  and the calculated position assuming that the 1981 transit is due to a 
  planet orbiting at 8~AU is intriguing. 
  Assuming that the planet that is detected on the image is the same as the object transiting in November 1981, 
  we estimate ranges of possible orbital distances and periods. 
  In the favored scenario, the planet orbits at $\sim$8~AU and was seen close to its quadrature position in the 2003 images. In this case, most of the uncertainties are related to error bars on the position in 2003. Uncertainties related to the stellar mass and orbital eccentricity are also discussed.
We find a semi-major axis in the range [7.6 - 8.7]~AU and an orbital period in the range [15.9 - 19.5]~years.
  We give predictions for imaging observations at quadrature in the southwest 
  branch of the disk in future years (2011-2015). 
  We also estimate possible dates for the next transits and anti-transits. 
%
}

   \keywords{Stars: planetary systems}

   \maketitle
%

\section{Introduction}
\label{Introduction}

When its IR excess was detected by the IRAS satellite in 1983, \bp\ became a star among the stars 
(see review by Vidal-Madjar et al.\ 1998, in which dust, gas, falling and orbiting evaporating small bodies, evidences of the presence of planets are discussed).
With the recent direct-imaging detection of a planet by Lagrange et al.\ (2009), a fascinating step in the study of that system has just been taken. The planet, hereafter named \bpicb , is detected on images obtained in 2003 at a projected distance of about 8~AU from the star, on the northeast side along the debris disk plane. Assuming the planet is close to quadrature, Lagrange et al.\ (2009) estimate that the orbital period should be about 16~years. Because the northeast side of the gas disk is receding (Olofsson et al.\ 2001; Brandeker et al.\ 2004), one can guess that if observed in 2003 in quadrature, the planet should have passed in front of the star in the years around 1999 and 1983.
This is an attractive coincidence with the possible observation of a planetary transit on November 10, 1981 (Lecavelier des Etangs et al.\ 1994, 1995) which might have been caused by the same planet.

In the photometric measurements obtained by Geneva observatory between 1975 and 1982, the light curve of \bp\ showed significant photometric variations in November 1981 with a transit-like event on November 10 (Lecavelier des Etangs et al.\ 1994, 1995). The probability that these variations are produced by statistical noise is less than $10^{-5}$ (Lecavelier des Etangs et al.\ 1995). 
These photometric variations were attributed to the transit of a giant comet (Lamers et al.\ 1997) or a planet orbiting at several AUs (Lecavelier des Etangs et al.\ 1994, 1995, 1997). From analysis of the transit light curve, Lecavelier des Etangs et al.\ (1997) obtained the following constraints on the possible transiting planet: 1) the period of the planet must be less than about 19~year to be compatible with measurements of November 10 and 11; 2) the radius of the transiting object must be 2.3 to 4.0~times the radius of Jupiter to be compatible with the transit ingress in the light curve. Both these constraints were made by assuming a nearly circular orbit for the planet. 
The stellar limb-darkening effect was also observed during that transit event. 
A slight color effect was detected with more absorption in the U~band, which may be explained by dust particles around the occulting planet with Rayleigh scattering, as recently observed in exoplanetary atmospheres (Lecavelier des Etangs et al.\ 2008a, 2008b). Material in the planet environment is indeed needed to explain the large occultation depth that is too much for a ``standard" planet alone. 
In light of the recent observations of transiting planets, the size inferred from the light curve is too large even for a single hydrogen-dominated warm and inflated gaseous 
planet in a young system. The observations are more consistent with a circumplanetary (proto-satellite) dust disk or a ring system around the planet, as recently inferred for the planet Fomalhaut\,b in a similar young debris disk (Kalas et al.\ 2008).

We subsequently did photometric surveys to search for another similar transit event. Unfortunately, 
we only obtained negative results. 
Nonetheless, this survey showed that all short periods below 1~year and most periods below 2 to 3~years are excluded (Nitschelm et al.\ 2000; Lecavelier des Etangs et al.\ 2005).

In summary, we have two detections: a planet observed at 8~AU from \bp\ in VLT images obtained in 2003 and an object that transited \bp\ in 1981. These two observations point toward objects with similar orbital periods and orbital phases. They could be two observations of the same object. The present paper evaluates 
this hypothesis and its consequences. 

In this paper we assume that the \bp\ stellar mass is 1.75$\pm$0.05\,$M_{\odot}$ (Crifo et al.\ 1997).
The distance to $\beta$\,Pic is given by Hipparcos measurements as $d$=19.3$\pm$0.2\,pc (Crifo et al.\ 1997).
In images obtained with the NaCo instrument at the Very Large Telescope (VLT), the point-like source is detected at 0.411$\pm$0.008 arcsec from the central star. Combined with the distance provided by Hipparcos, this gives a projected distance of 7.93$\pm$0.24\,AU. 

\section{Is \bpicb\ a transiting planet ?}

\begin{figure}[tb]
\psfig{file=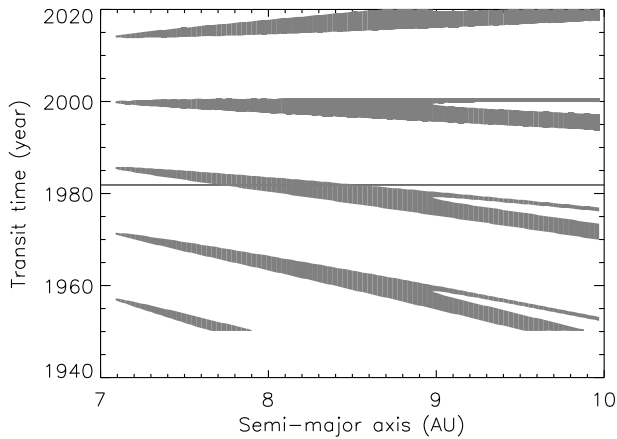,width=\columnwidth,angle=90} 
\vspace{0.4cm}
\caption[]{
Plot of the transit date for \bpicb\ assuming that it is a transiting planet (gray zones). Uncertainty on the \bpicb\ position in November 2003 is taken into account.  
The eccentricity is assumed to be less than or equal to 0.1. Above 9\,AU, the solutions are doubled because of the degeneracy of the observed position of \bpicb\ in 2003, either before or after the quadrature. 
The horizontal line shows the date of the 1981 photometric event. 
\label{transit_a}}
\end{figure}

The \bp\ disk is constrained to be inclined by only a few degrees 
(Lecavelier des Etangs et al.\ 1993; Kalas et al.\ 1995; Heap et al.\ 2000; 
Golimowski et al.\ 2006). If a planet is located within 
the disk, the transit probability is enhanced by this favorable configuration. 
Since \bpicb\ is apparently within the disk, we can assume that it could 
be a transiting planet. Under this hypothesis, we can estimate when the transit occurred in the past.

Using the observed projected distance of \bpicb\ (and its associated error bars) 
and assuming that it is a transiting planet, we estimated the date of transits 
for an eccentricity range of 0 to 0.1 (Fig.~\ref{transit_a}). 
For a semi-major axis between 7 and 10~AU, we find a wide range 
of solutions consistent with the observation of a transit in 1981. 

We conclude that \bpicb\ could be a transiting planet, and, in this case, 
the similitude of \bpicb\ with the transiting object of 1981 is likely. 
In the following, we assume that the 
imaged planet is the same as the transiting object in November 1981. 
With this assumption, we obtain new constraints on the possible 
orbital characteristics of \bpicb\ (Sect~\ref{as the transiting planet}), 
and we make predictions for the forthcoming observations 
(Sect.~\ref{Future observations}). Comparison of these predictions 
with observations will allow to disprove or improve the present model. 

\section{\bpicb\ as the transiting planet of 1981}
\label{as the transiting planet}

Assuming that \bpicb\ is the transiting object of 1981, we have two measurements
to constrain the orbit characteristics. In this section, we evaluate its
semi-major axis and orbital period. Uncertainties on the planet position 
and on the stellar mass, as well as eccentric orbits are also considered. 

\subsection{Semi-major axis and orbital period}
\label{Semi-major axis and Orbital period}

\begin{figure}[tb]
\psfig{file=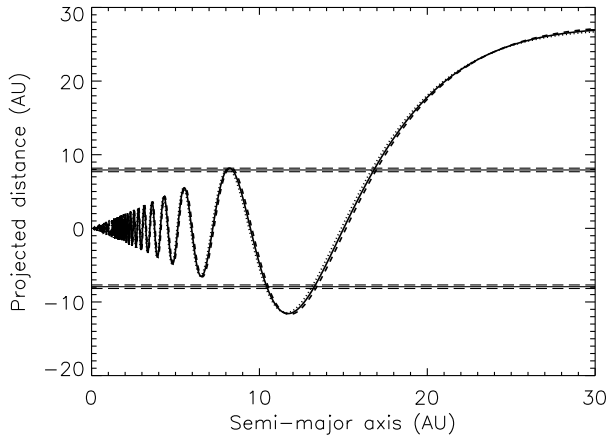,width=\columnwidth,angle=90} 
\vspace{0.4cm}
\caption[]{Position (projected distance to the star) of a planet orbiting 
around \bp\ as a function of its semi-major axis, assuming that the planet transited in November 10, 1981. The position is given for the observation date of the discovery images of \bpicb\ (November 10, 2003). 
The orbit is supposed to be circular. The stellar mass is 1.75\,$M_\odot$.
The horizontal lines represent the measured position of \bpicb\ in 2003 (horizontal solid lines) with
associated error bars (horizontal dashed lines). The positive positions are for planets moving away from the observer, and negative positions for planets moving toward the observer. If \bpicb\ orbits in the same direction as the gaseous disk, the measured position in 2003, in the northeast branch of the disk, corresponds to a positive measurement at +7.93$\pm$0.24\,AU. 
It is noteworthy that, if a planet orbiting at 8\,AU transited in 1981, then it must be exactly
at the position of \bpicb\ in 2003. 
\label{d_a}}
\end{figure}

\begin{figure}[tb]
\psfig{file=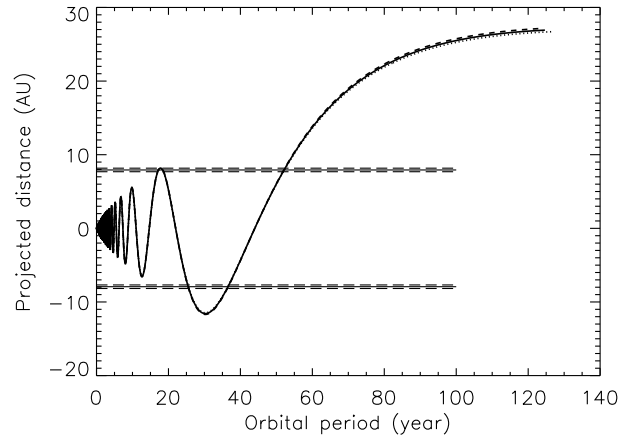,width=\columnwidth,angle=90} 
\vspace{0.4cm}
\caption[]{
Same as previous plot for the position of the planet as a function of its orbital period. 
\label{d_p}}
\end{figure}

We calculated the position of a planet as a function of semi-major axis and orbital period assuming that it transited in November 10, 1981 (Figs.~\ref{d_a} and~\ref{d_p}). 
The planet position was calculated for the date of VLT-images observations in 2003. 
We see that the planet position is consistent with the observed \bpicb\ position in a few cases. 

The first possible solution is a planet orbiting at $\sim$8\,AU and observed 
closed to its quadrature position in 2003. In this case the planet is observed when
moving away from the observer. It is most intriguing that this is exactly the 
disk rotation obtained from gaseous emission lines observations 
(Olofsson et al.\ 2001; Brandeker et al.\ 2004). 
In other words, if the planet transiting in 1981 has an $\sim$8\,AU semi-major axis, it must
be observed at the position of \bpicb\ in 2003. This intriguing coincidence 
justifies the hypothesis made in this paper that the two objects are in fact the same 
object. In the case of a planet orbiting at 8\,AU, the orbital period is 17.0 to 18.6~years 
(Fig.~\ref{d_p}).

If \bpicb\ has been observed when moving toward the observer, 
there are two solutions for the semi-major axis : 
$\sim$10.5 and $\sim$13.5~AU (Fig.~\ref{d_a}). 
However, although planets have been observed with very high orbital 
inclination ({\it e.g.}, H\'ebrard et al.\ 2008), 
it seems unlikely that a planet could have a retrograde motion relative to the 
circumstellar disk. These two solutions are therefore considered as unlikely. 

The last solution is a semi-major axis of 16.8~AU, in which case the period is about 52~years long. In this case, the planet has moved along a little less than half an
orbit from 1981 to 2003, and has been observed in 2003 at a projected distance of 8~AU just before the opposition (or ``secondary transit").
In this solution, the planet was at greater projected distance from \bp\ before 2003. For instance, in 1995, the planet would have been at 0.9 arcsec from \bp\ and may have been detected in HST or adaptative optics ground-based observations. Therefore, this solution also appears less likely than the first one. 

In conclusion, we note the very intriguing coincidence of the 2003 position of \bpicb\ with the
calculated position of the 1981 planet if it orbits at $\sim$8 AU. Other scenarios appear less
likely. In the following we consider the planet orbiting at 8~AU and observed close 
to quadrature in 2003 as the favored solution. 
 
\subsection{Uncertainties}
\label{Uncertainties}

\begin{figure}[tb]
\psfig{file=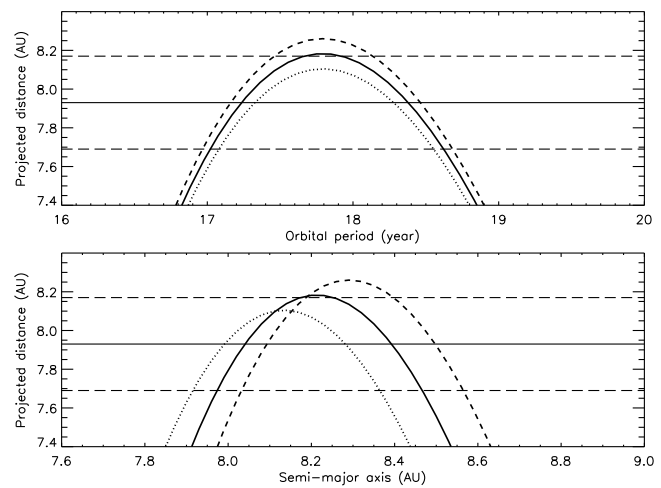,width=\columnwidth,angle=90} 
\vspace{0.4cm}
\caption[]{
Zoom of the two previous plots for the region of semi-major axis around $\sim$8~AU (lower panel) 
and orbital period around $\sim$18~years (upper panel).
The horizontal lines represent the position of \bpicb\ in 2003 (solid line) with
associated error bars (dashed lines). 
The projected distance is given for stellar masses of 1.75\,$M_\odot$ (thick solid line), 
1.7\,$M_\odot$ (thick dotted line), and 1.8\,$M_\odot$ (thick dashed line). 
\label{d_a_p_zoom}}
\end{figure}

We can evaluate the uncertainties on the period and the semi-major axis, 
due to the uncertainties on the stellar mass and on measured position in 2003.

We first consider the favored solution and we plot the positions 
of the planet in November 2003 for various stellar masses (Fig.~\ref{d_a_p_zoom}). 
With a stellar mass of 1.75\,$M_\odot$, the calculated positions 
are consistent with the observations for semi-major axis between 8.0~AU 
and 8.5~AU and for periods between 17.0~years and 18.6~years.
Including the uncertainty on the stellar mass, 
we obtain semi-major axis between 7.9~AU and 8.6~AU, and periods between 
17.0~years and 18.6~years.
We conclude that the uncertainty on the stellar mass has no significant 
impact on the period and a low impact on the semi-major axis estimation. 
Most of the uncertainties come from the uncertainties on the position 
of \bpicb\ in 2003 images. 

\subsection{Eccentric orbit}

\begin{figure}[tb]
\psfig{file=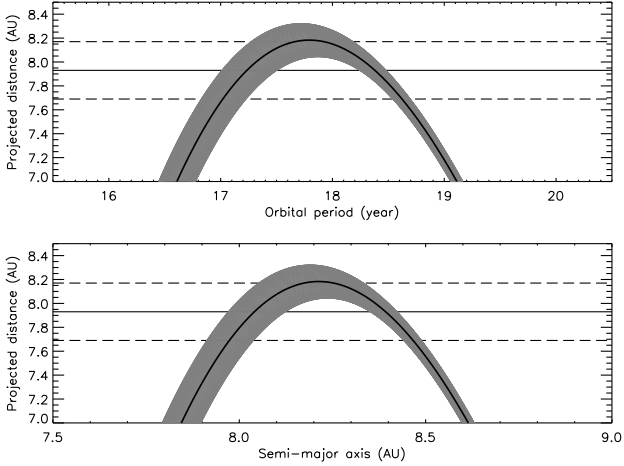,width=\columnwidth,angle=90} 
\vspace{0.4cm}
\caption[]{
Same as previous plot for eccentricity between 0 and 0.02. All possible longitudes of periastron and  
all eccentricities in the range [0, 0.02] are taken into account. The family of solutions 
is plotted by the grey zone. 
\label{d_a_p_e002}}
\end{figure}

\begin{figure}[tb]
\psfig{file=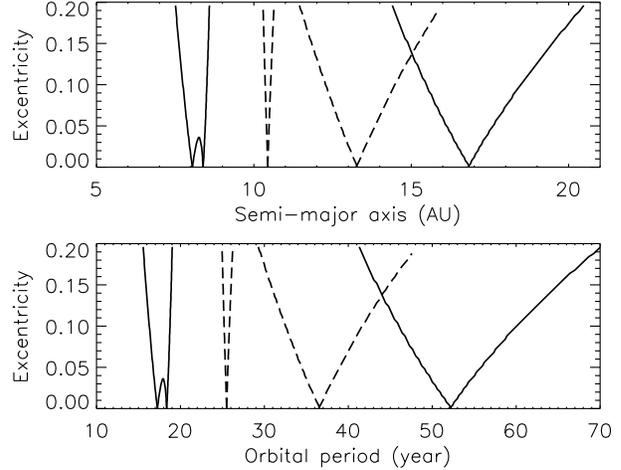,width=\columnwidth,angle=90} 
\vspace{0.4cm}
\caption[]{
Plot of possible solutions for the semi-major axis and the orbital period of a planet transiting 
in November 1981 and located at the position of \bpicb\ in 2003. The ranges of possible solutions are
given as a function of eccentricity of the planet orbit, and neglecting the uncertainty on the position 
of \bpicb. 
The ranges are given for planets observed moving away from the observer in 2003 (solid line) and moving toward the observer in 2003 (dashed line). 
Because \bpicb\ has been observed in the side of the gaseous disk moving away from the observer, 
the first case is favored. 
For low eccentricity and a planet at $\sim$8\,AU, there are two solutions for a planet observed in 2003 : a planet observed just before or just after the quadrature. 
\label{a_p_e}}
\end{figure}

Unfortunately, the knowledge of 2 positions on the orbit (projected distance to the star, and time of transit) is not enough to fully constrain
the orbit. The first limitation is the degeneracy of the projected distance in two positions on the orbit, before and after the quadrature. Moreover, the knowledge of two positions does not allow the eccentricity of the orbit to be constrained.
In the previous sections (Sects.~\ref{Semi-major axis and Orbital period} and ~\ref{Uncertainties}), we assumed circular orbits.

Some plausible guess can be made on the eccentricity of the orbit.
Indeed, even before the discovery image of \bpicb, the presence of a planet was needed to explain
all the phenomenons observed in the \bp\ disk (Vidal-Madjar et al.\ 1998).
For instance, a massive planet is suspected to be responsible for the peculiar orbits
of the cometary bodies also called Falling Evaporating Bodies
(FEBs; Beust et al.\ 1990, 1991; Vidal-Madjar et al.\ 1994; Beust \& Morbidelli 1996, 2000),
and a planet can easily explain the observed assymetries and warp of the dust disk
(Lecavelier des Etangs et al.\ 1996, 1998; Mouillet et al.\ 1997; Augereau et al.\ 2001).
The model of Beust \& Morbidelli (1996, 2000) gives detailed explanation of the onset of the FEBs phenomenon
and all its observational characteristics by mean motion resonances with a massive planet.
They concluded that a planet of about 1 $M_{\rm Jup}$ at about 10~AU with a moderate eccentricity
of e$\approx$0.05 can explain both the warp of the dust disk and the FEBs phenomenon.
More recently, belt structures at about 6.4, 16, 32, and 52~AU were observed (Wahhaj et al.\ 2003; Okamoto et al.\ 2004; Telesco et al.\ 2005; Golimowski et al.\ 2006).
Reviewing the evidence of planets in the \bp\ disk and using dynamical simulations, Freistetter et al.\ (2007)
obtained additional constraints on these planets. They found that two low mass planets ($<0.6$ and $<0.2\,M_{\rm Jup}$,
respectively) at about 25~AU and 45~AU can explain the observed outer belts. More importantly here,
they also found that a planet at about 12~AU with a 2 to 5 Jupiter mass and an eccentricity $e\le 0.1$
is able to account for the main warp, the two inner belts, and the FEBs. In short, the large set of
phenomenons observed in the \bp\ disk points toward the presence of a massive planet
at $\sim$10\,AU with a moderate eccentricity of less than about 0.1.

Assuming an eccentricity of 0.02, we calculated the range of solutions for the planet's
semi-major axis and period (Fig.~\ref{d_a_p_e002}).
Considering uncertainties on the position measured in 2003, because of the 
eccentricity, uncertainties on the semi-major axis and period
are slightly increased by 
about (+0.1/-0.2)\,AU and (+0.02/-0.05)\,years, respectively.

For various eccentricities, we plotted the range of
solutions for the semi-major axis and the period (Fig.~\ref{a_p_e}).  
Here, to evaluate the uncertainty only due to the eccentricity,
we do not include the uncertainty on the planet position when observed 
in November 2003.
For eccentricity less than or equal to 0.1, in the first range of solution
the semi-major axis is between 7.7~AU and 8.5~AU, and the orbital period is between
16.2~years and 18.8~years. This corresponds to an increase in uncertainties by
(+0.1/-0.3)\,AU and (+0.4/-1.0)\,years, respectively.

We therefore conclude that the uncertainty introduced by eccentricity mainly allows for a
slightly smaller semi-major axis and correspondingly shorter periods. For moderate eccentricity,
this uncertainty remains lower than the uncertainty due to the planet position measurement.

Finally, for the favored scenario, including all uncertainties 
and assuming an eccentricity $e\le 0.1$, we find a 
range of semi-major axis and orbital periods of [7.6 - 8.7]~AU 
and [15.9 - 19.5]~years, respectively.

\section{Future observations}
\label{Future observations}

\begin{figure}[tb]
\psfig{file=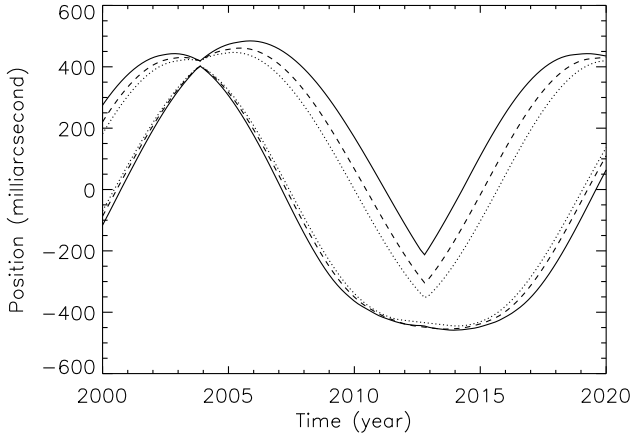,width=\columnwidth,angle=90} 
\vspace{0.4cm}
\caption[]{
Envelope of possible positions of \bpicb\ as a function of time, assuming that the object transited in November 1981. The position is given in distance from \bp\ in milliarcsec, positive values corresponding to the northeast branch of the disk, where \bpicb\ has been observed. 
The envelope is given for eccentricities of $e=$0.02 (dotted lines), $e=$0.05 (dashed lines), and $e=$0.1 (solid line). The stellar mass is assumed to be 1.75\,$M_{\odot}$. Plots with stellar mass of 1.7 and 1.8\,$M_{\odot}$ give very similar results. 
\label{Pos_t_2000_2020}}
\end{figure}

Assuming that the transit object of 1981 is 
the \bpicb\ planet discovered in 2003 images, we obtained constraints on the possible 
orbital solution of this planet. In a next step, using these results, we can provide 
predictions for future observations. These predictions can be used in three ways: first, to 
confirm or infirm the hypothesis made here by comparing with the result of future observations; 
second, to schedule observations and, in case the hypothesis is correct, to increase the 
likelihood of observing \bpicb\ again ; finally, to understand which observations 
will allow several possible scenarios to be distinguished. 

\subsection{Five years later, where is \bpicb ?}

The question of the true nature of the candidate planet \bpicb\ remains, because it has
been observed only once. New observations are required to disprove the possibility of 
a background object (although this is extremely unlikely; Lagrange et al.\ 2009).
Since \bpicb\ was discovered in 2003 images, we need to compare the predicted positions 
in 2009-2010 for a background object and for a planet. 

In Fig.~\ref{Pos_t_2000_2020} we plotted the possible positions of \bpicb\ assuming that 
it transited in November 1981 and that it has an $\sim$18~year period. In this case, 
we can see that in 2009 the planet is most likely close to the opposition (secondary transit). 
We find that in 2009 the planet will be 
within 300~milliarcsecond from the central star. This is certainly the less favorable 
period for a new detection. In addition that, in the case of an orbital period 
of $\sim$52~years (although unlikely), the planet will also be near opposition in 2009. 

Furthermore, \bp\ has a proper motion of about 80~milliarcsecond per year in a direction 
close to the direction of \bpicb\ as observed in 2003. Unfortunately, 
this motion will locate any background object at the position of \bpicb\ in 2003 
very close to \bp\ in 2009, at less than 200~milliarcsecond from the star
(Ehrenreich, private communication).

Therefore, for the three possible scenarios (background object, 
planet at 8~AU, or planet at $\sim$17~AU), in most cases \bpicb\ is predicted to be very close to \bp\ in the sky in 2009. With the high star-planet contrast and a distance of less than $\sim$300 milliarcsec from \bp , 
new images obtained in a near future
are unlikely to allow differentiation of these three scenarios.
Alternatively, if the object is seen at a different location, this
would invalidate all these three possibilities.  

\subsection{Schedule of forthcoming quadrature}
\label{Schedule of forthcoming quadrature}

In the favored scenario in which the planet has an orbital period of $\sim$18~years and a semi-major 
axis of $\sim$8~AU, the next quadrature position will be reached between 2011 and 2015 (Fig.~\ref{Pos_t_2000_2020}). 
We predict that in 2013, \bpicb\ should be observed between 200 and 500 milliarcsecond from \bp\ in the southwest branch of the disk. If future observations disprove this prediction, this would imply that the hypothesis that \bpicb\ transited in November 1981 is wrong.

\subsection{Schedule of forthcoming transits and anti-transits}

For extrasolar planets studies, the transit is a key configuration
for characterizing the orbit 
({\it e.g.} H\'ebrard \& Lecavelier des Etangs 2006), 
to search for the signature of evaporating bodies 
({\it e.g.} Lecavelier des Etangs et al.\ 1997b, 1999a, 1999b)
or to search for atmospheric signatures 
({\it e.g.} Vidal-Madjar et al.\ 2003, 2004, 2008; Ehrenreich et al.\ 2006, 2007). 

If \bpicb\ did transit in 1981, it would be of prime interest to know when the next transits 
and anti-transits will happen. 
Assuming that the planet orbits at $\sim$\,8 AU, with an eccentricity of less than or equal to 0.1 and 
with a stellar mass of 1.75$\pm$0.05\,$M_{\odot}$, we find that in the past the planet transited between October 1997 and
May 2001 and that the following anti-transit should occur between October 2005 and February 2011. 
Unfortunately, although it would be wonderful if this anti-transit were observed with Spitzer, the uncertainty precludes for scheduling this kind of observation. 
The following primary transit should happen between September 2013 and December 2020. 

Even if this primary transit is preceded by light variations over a few days (Lecavelier et al.\ 1995), it 
is extremely difficult to anticipate this kind of observation. It is more likely that new image
observations in the coming quadrature (Sect.~\ref{Schedule of forthcoming quadrature}) 
could help to better constrain the time of the next transit. 

\section{Conclusion}

We have constrained the possibilities for future observations
of \bpicb\ assuming that it was the transiting object of 1981. After acknowledging the uncommon potentialities of this transiting planet, we are now waiting for 
new data to improve or refute the proposed scenario. 

Presently, continuing or starting new photometric surveys of \bp\ will be useful.  
For instance, long time-scale variations by the occulting belt of dust in 1:1 resonance with the 
planet (Lecavelier des Etangs et al.\ 1997) could be used to foresee the transit event. 
Indeed dust accumulated close to the Lagrange point should be responsible for
some variation in the extinction when moving in front of the star. 
The highest extinction should be reached about 3 years before the transit, 
that is, between $\sim$2011 and $\sim$2016. 
If these long time-scale variations are detected, they could be used to 
further constrain the possible time of the next transit.  

If future observations happen to confirm that \bpicb\ is a transiting planet, this planet would be an extraordinary transiting planet, because it is 
\begin{enumerate}
\item a planet transiting in front of a 3.8 magnitude star. 
By comparison with what has been done in the case of planets transiting 
7$^{\rm th}$ magnitude stars ({\it e.g.} Charbonneau et al. 2002; Sing et al.\ 2008a, 2008b), 
the atmosphere of this planet could be probed with unprecedented detail 
for an extrasolar planet. 
\item a young planet with circumplanetary material. Through detailed transit observations,  
this planet could give unique information on the planet environment including rings and satellites, 
at a stage when satellites are still forming or just formed.
\item a planet at 8~AU. This would give access to transit observations of a planet far from its parent star, a situation more like the giant planet of the solar system and different from the hot-Jupiters, presently the only known transiting planets. 
\end{enumerate}
 
As a conclusion, we are now waiting for new observations to confirm or disprove 
the hypothesis developed in this present work. If confirmed, \bpicb\ could soon become a mine of information on young extrasolar planets. 

\begin{acknowledgements}
We warmly thank C.~Nitschelm who first pointed out the photometric measurements of \bp\ obtained by the Geneva Observatory. We thank D.~Ehrenreich, R.~Ferlet, and A.-M.~Lagrange for their comment on the manuscript.
We are also grateful to Drs. F.~Bouchy, G.~Burki, M.~Deleuil, J.-M.~D\'esert, G.~H\'ebrard, H.~Lamers, G.~Perrin, and D.~Sing for fruitful discussions on the subject of the present article. 
\end{acknowledgements}

\end{document}